\documentstyle[aps,twocolumn,cite,prb,epsfig,amstex]{revtex}

\begin{document}

\title{Superconducting Proximity Effect and\\Universal Conductance
Fluctuations}
\author{Tero T. Heikkil\"a and Martti M. Salomaa}

\address{Materials Physics Laboratory, Helsinki University
of Technology,\\P.O. Box 2200 (Technical Physics), FIN-02015 HUT,
Finland
}

\author{Colin J. Lambert}

\address{Department of Physics, Lancaster University, Lancaster
LA1 4YB, UK}

\date{\today}

\maketitle
\begin{abstract}
We examine universal conductance fluctuations (UCFs) in mesoscopic
normal-superconducting-normal (N-S-N) structures using a numerical
solution of the Bogoliubov-de Gennes equation. We discuss two cases
depending on the presence (``open'' structure) or absence (``closed''
structure) of quasiparticle transmission.
In contrast to N-S structures, where the onset of superconductivity
increases fluctuations, we find that UCFs are suppressed by
superconductivity for N-S-N structures. We demonstrate that the
fluctuations in ``open'' and ``closed'' structures exhibit distinct
responses to an applied magnetic field and to an imposed phase
variation of the superconducting order parameter.
\end{abstract}

\pacs{74.50.+r,74.40.+k,74.80.-g}

\narrowtext
Transport properties of mesoscopic structures in contact with
superconducting segments have attracted a great deal of
interest~\cite{lambert98}. In these structures,
the superconducting proximity effect induces novel transport
phenomena, including sub-gap anomalies in normal-superconducting (N-S)
contacts and phase-dependent conductances in Andreev
interferometers. Quasiclassical theory, which has been widely applied,
has been successful in explaining phenomena associated with
ensemble-averaged properties but it cannot describe fluctuations about
the mean~\cite{lee87}.

In normal diffusive nanostructures of size smaller than the phase
coherence length $l_\phi$ but larger than the elastic mean free path
$l_{\text{el}}$, the conductance fluctuates by an amount of order
$e^2/h$. These fluctuations can be observed by varying either the
magnetic field, gate potential, or impurity configuration. The
fluctuations are universal within the diffusive regime: their
magnitude is partly determined by symmetry and does not depend on
the degree of disorder or system size. Such a symmetry-breaking
operation is the application of an external magnetic field which
breaks time reversal (TR) symmetry and decreases the rms fluctuations
by a factor of $\sqrt{2}$. 

For normal-superconducting systems, it has been
shown~\cite{brouwer95,altland97,hecker97} that the onset of
superconductivity increases the rms magnitude of UCFs by a factor which
depends on the underlying order parameter. Beenakker and Brouwer
(BB)~\cite{brouwer95} predict that in the absence of phase gradients
for the order parameter, breaking TR symmetry has a negligible effect
on the rms conductance fluctuations $\delta G_{\text{NS}}$ of a N-S
structure. Altland and Zirnbauer (AZ)~\cite{altland97}, on the other
hand, find that $\delta G_{\text{NS}}$ decreases by a factor of $\sqrt
2$, provided that the average phase change of an Andreev-reflected
quasiparticle is zero. When the latter condition is satisfied, AZ
predict that in zero magnetic field, 
\begin{equation}
\delta G_{\text{NS}}(B=0)=2\sqrt{2} \delta G_{\text{NN}}(B=0), 
\label{eq:aznomagn}
\end{equation}
where $\delta G_{\text{NN}}$ is the rms conductance fluctuation when
the S contact is in the normal state. Since switching on a magnetic
field decreases both $\delta G_{\text{NS}}$ and $\delta G_{\text{NN}}$
by a factor of $\sqrt 2$, they also find
\begin{equation}
\delta G_{\text{NS}}(B\ne 0)=2\sqrt{2} \delta G_{\text{NN}}(B\ne 0)=
2\delta G_{\text{NN}}(B=0).
\label{eq:azmagn}
\end{equation}
In contrast, since BB predict that $\delta G_{\text{NS}}$ is almost
insensitive to TR symmetry breaking, they obtain
\begin{gather}
\delta G_{\text{NS}}(B=0)\approx 2 \delta G_{\text{NN}}(B=0)\\
\delta G_{\text{NS}}(B\ne 0)= 2\sqrt{2} \delta
G_{\text{NN}}(B\ne 0) = 2\delta G_{\text{NN}}(B= 0).
\label{eq:bbresults}
\end{gather}
Thus the AZ and BB scenarios yield essentially the same fluctuations
for finite $B$, a result which has recently been tested
experimentally~\cite{hecker97}, whereas they yield fluctuations of 
different magnitudes in zero field. 

In this paper we examine for the first time the crossover between
the AZ and BB scenarios and predict that this crossover is
observable in closed Andreev interferometers. The BB and AZ
calculations are applicable to N-S structures with a single N contact,
whereas many experiments involve two or more N contacts
\cite{hartog}. We also examine the field and phase dependence of
fluctuations in the conductance $G_{\text{NSN}}$ of N-S-N structures
possessing two normal (N) contacts
\cite{takane91,altshulerbrouwer}. For the considered geometries, these
have not been explored before in the literature. 

We consider the two limiting cases of N-S-N structures shown in
Fig.~1, comprising two N-reservoirs attached to a normal diffusive
region with superconducting contacts and a current flowing from left
to right. In Fig.~1a, the two superconductors do not intersect the
diffusive normal region and affect transport only through the proximity
effect. In this structure, quasiparticle transmission between the
reservoirs is possible and therefore we refer to this as ``open''.
In the second, ``closed'', structure of Fig.~1b, a superconductor
of length $L_{\text{S}}$ divides the diffusive region into two normal 
regions, each of length $L_{\text{N}}$. The length $L_{\text{S}}$ of
the superconducting segment is much greater than the superconducting
coherence length $\xi$ and therefore quasiparticle transmission
through the superconductor is suppressed. 

\begin{figure}[h]
\centering
\setlength{\unitlength}{1cm}
\epsfig{file=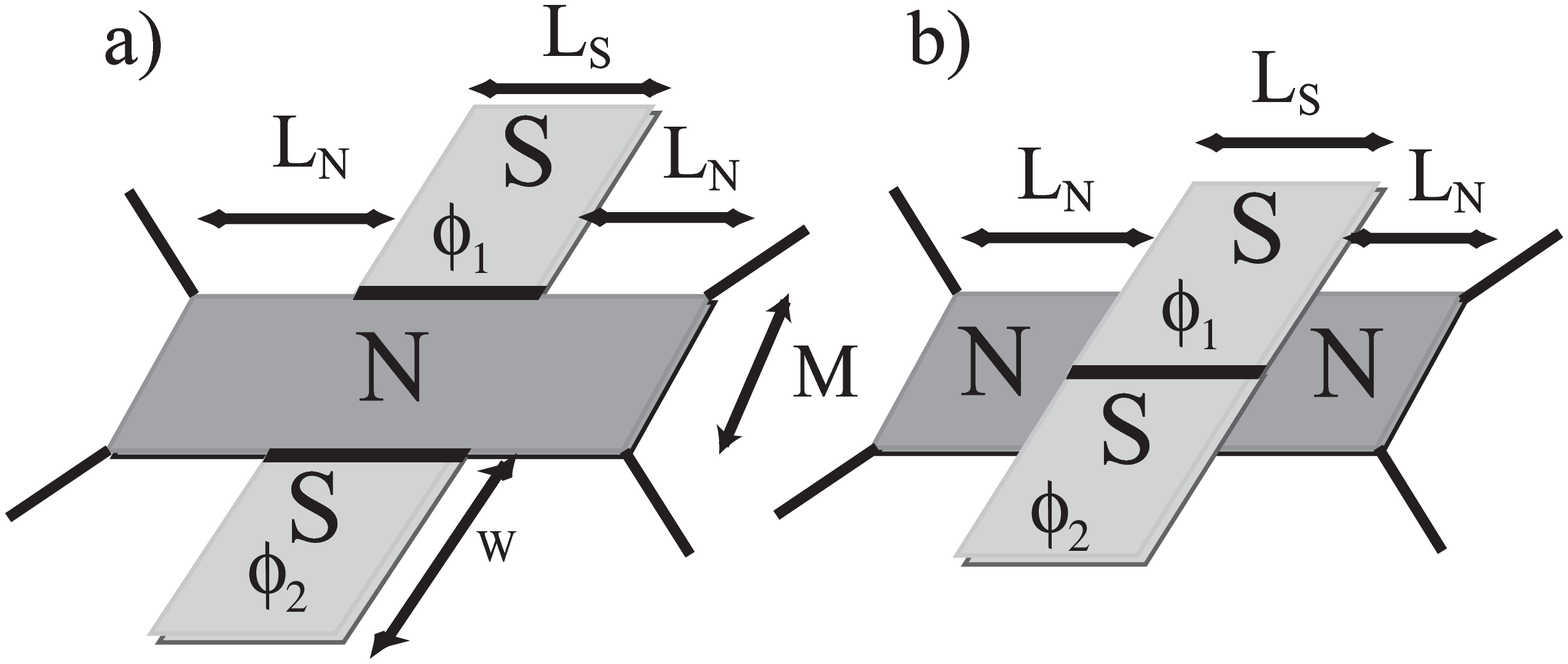,width=8.26cm}
\label{fig:structures}
\end{figure}

{\small FIG. 1. a) An open Andreev interferometer where the
superconducting segments are located outside the classical path of the
current. The N-S interfaces contain an optional Schottky barrier with
transmissivity $\Gamma$ (i.e., transmission probability per
channel). The S segments have an order-parameter phase difference of
$\phi=\phi_1-\phi_2$ b) A closed interferometer where the
superconducting segments divide the normal diffusive region into two
halves with no transmission from left to right. At the interface
between the S segments there is a tunnel barrier of width 1 site and
transmissivity $\Gamma=0$.}   

\smallskip

In the linear-response limit and at zero temperature, the relationship
between the  current $I$ and the reservoir potential difference
$v_1-v_2$ was first obtained in ~\cite{lambert91} and the
corresponding conductance (in units of $2e^2/h$) may be written in the
form 
~\cite{lambert98}, 
\begin{equation}
G_{\text{NSN}}=
T_0+T_a +2\left[\frac{R_aR_a'-T_aT_a'}{R_a+R_a'+T_a+T_a'}\right],
\label{eq:lambertformula}
\end{equation}
where unprimed [primed] quantities $T_0$ and $R_0$ ($T_a$ and $R_a$)
are normal (Andreev) transmission and reflection probabilities for
quasiparticles from lead 1 [lead 2].
To evaluate this expression, we solve the Bogoliubov-de Gennes
equation numerically on a two-dimensional tight-binding lattice with
diagonal disorder, utilizing the decimation method to evaluate the
transport coefficients of a two-probe structure~\cite{lambert94}. The
magnetic field is introduced via the Peierls substitution and the
disorder in the normal diffusive area is introduced by varying the
site energies at random within the range $[-w/2,w/2]$. 

Conductance fluctuations are universal only if transport through the
normal structures is diffusive. Therefore, following
~\cite{takane}, we choose the disorder $w$ such that the
conductivity of the diffusive normal region is independent of its
length $L$. For a sample of width $M=35$, in the absence of
superconductivity, Fig.~2 shows the dimensionless
conductivity $ l=\langle G \rangle L/M$ as a function of $L$ for four 
different values of disorder $w$. In the quasi-ballistic regime,
$\langle G \rangle$ approaches the Sharvin conductance and $l$
increases linearly with $L$, whereas at intermediate $L$, $l$ exhibits
a plateau, corresponding to the diffusive regime of interest. In units
of the lattice constant, this plateau value of $l$ is equal to the
elastic mean free path $l_{\text el}$. Therefore, Fig.~2 represents a
mapping between the model-dependent parameter $w$ and the
experimentally accessible quantity $l_{\text el}$. Finally, for values
of $L$ larger than the localization length $\lambda \approx N
l_{\text{el}}$, where $N=30$ is the number of open channels, $l$
decreases with increasing $L$. In Figs.~4--5 we choose $w=1.5$,
corresponding to a mean free path of $l_{\text{el}} \approx 8$. The
other  length scales (in units of the lattice constant) are $M=35$,
$L_{\text{N}}=75$, $L_{\text{S}}=40$, $W=10$ and $\xi=10$, where $\xi$
is the superconducting coherence length setting the scale for the
decay of zero-energy wavefunctions within the superconductor
\cite{checknote}.  

\begin{figure}[h]
\centering
\epsfig{file=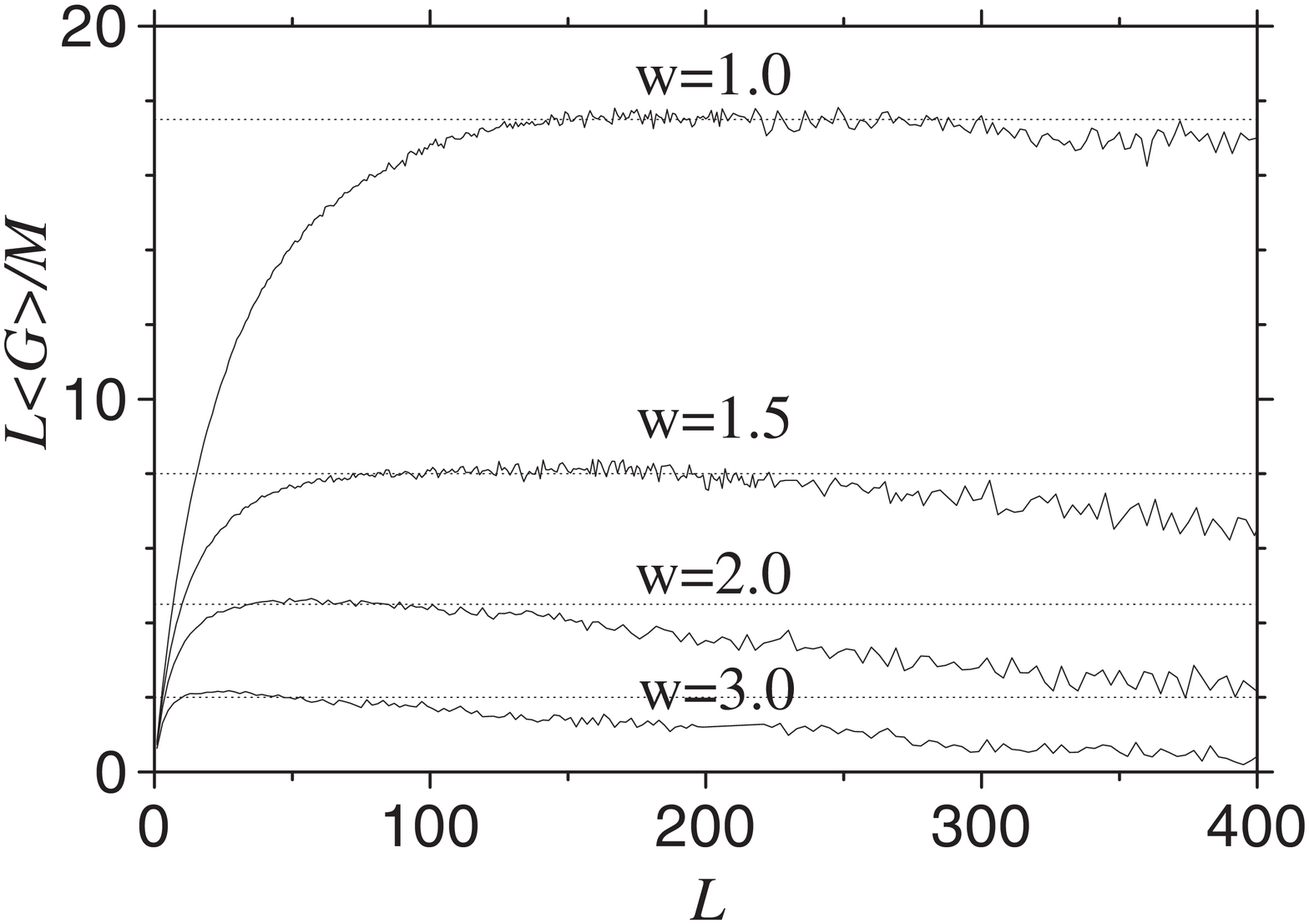,width=8.26cm}
\label{fig:mfp}
\end{figure}

{\small FIG. 2. Quantity $\langle G \rangle L/M$ plotted as function of $L$
for different strengths $w$ of disorder. The region where the curve is
essentially constant corresponds to diffusive transport. The constant
defines the mean free path $l_{\text{el}}$. The dotted lines at
$l=l_{\text{el}}$ are guides to the eye.}

\smallskip

Figure 3 shows the rms fluctuations as a function
of $l$ for a normal structure in the absence of superconductivity and
for the N-S-N structures of Fig.~1 (with zero magnetic field and
a constant order-parameter phase); it demonstrates that fluctuations
are independent of $l$. This universal behavior is in agreement with
previous numerical \cite{marmorkos93} and analytical
\cite{takane91,beenakker93} works. In the diffusive regime, the
presence of superconductivity for both structures in Fig.~1 suppresses
the fluctuations by a factor of order $\sqrt{2}$. 

For the closed structure in Fig.~1b, all transmission coefficients
necessarily vanish, and for structures of type 1a, the presence of
disorder and a proximity-induced suppression of the density of states
may also cause transmission coefficients to be negligible. In this
case, Eq.~(\ref{eq:lambertformula}) reduces to a sum of two N-S 
resistances, $G_{\text{NSN}}=1/(2R_a) + 1/(2R'_a)$, associated with
Andreev reflection of quasiparticles from the two decoupled
reservoirs. By adding two statistically independent N-S resistances in
series, one finds that in the absence of transmission, the
fluctuations $\delta G_{\text{NSN}}$ of the total conductance are
related to the fluctuations $\delta G_{\text{NS}}$ of the
N-S-conductance by $\delta G_{\text{NSN}} = \delta
G_{\text{NS}}/2\sqrt{2}$. Therefore, in the absence of transmission, the
$\sqrt 2$ decrease in $\delta G_{\text{NSN}}$ due to the onset of
superconductivity is a consequence of this relation and Eq.~(3)
\cite{transmissionnote}.   

\begin{figure}[h]
\centering
\epsfig{file=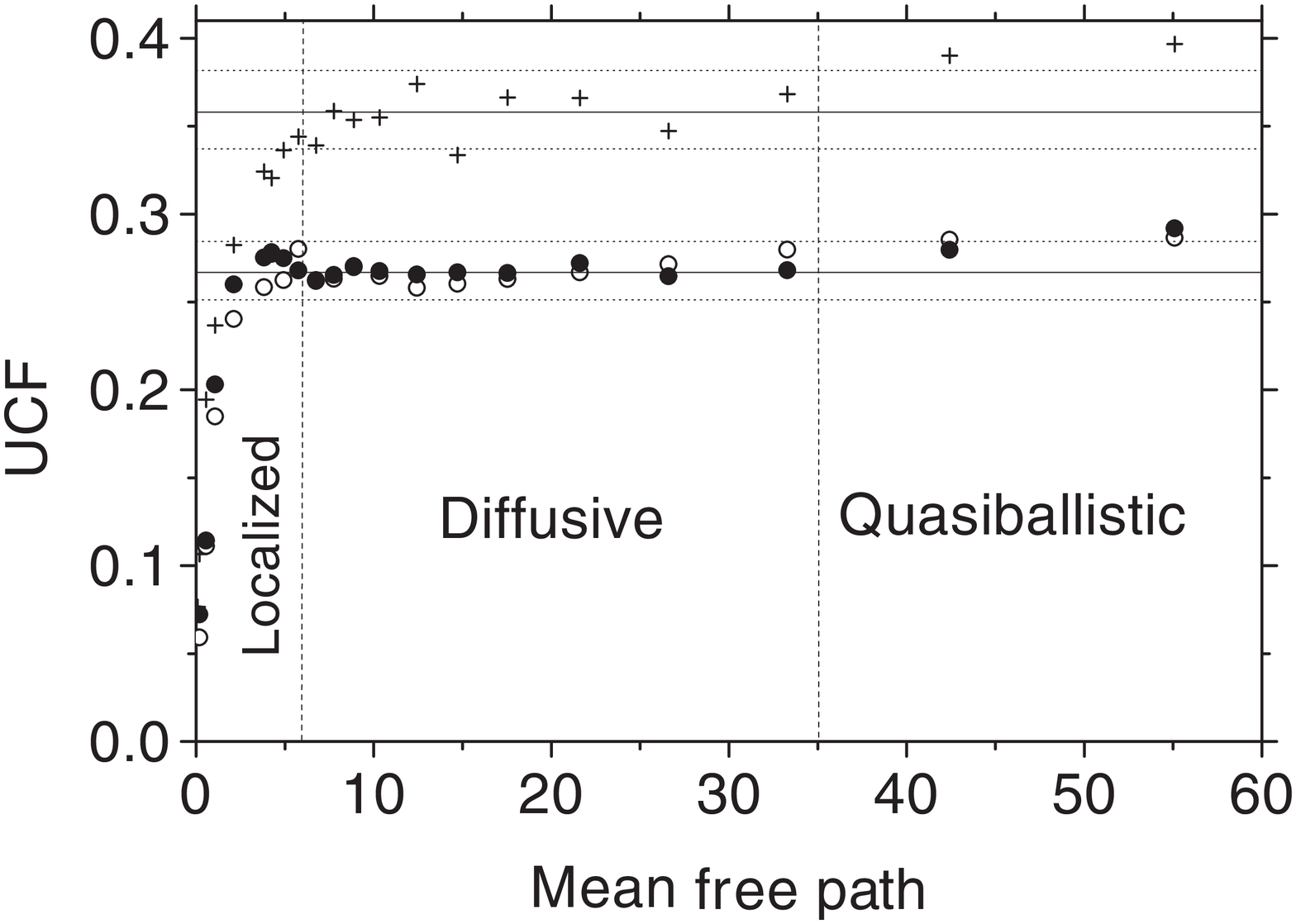,width=8.26cm}
\setlength{\unitlength}{1cm}
\label{fig:universality}
\end{figure}

{\small FIG. 3. Universal conductance fluctuations (rms deviations) in units
of $2e^2/h$ as function of $l$ for a normal structure (crosses), open
N-S-N structure (open circles), and closed N-S-N structure (closed
circles), obtained from over 500 realizations of disorder. The mean
free path extends over three regimes: localized ($l<6$), diffusive
($6<l<35$) and quasiballistic ($l>35$). The solid lines at $\delta
G=\delta G_{\text{N}}=0.358$ and $\delta G=\delta G_{\text{S}}=0.267$
show the average of the standard deviations within the diffusive
regime in the normal and superconducting cases, respectively. The
dashed lines indicate 95\% confidence intervals.}

\smallskip

Figure 4a shows that in the presence of a large enough magnetic field,
the rms fluctuations of both a normal structure and the open structure
in Fig.~1a are suppressed by a factor of order $\sqrt{2}$. However, in
agreement with the prediction of BB, the suppression of $\delta
G_{\text{NSN}}$ for the closed structure of Fig.~1b is at most of the
order of 10\%.  

To demonstrate that a crossover from the BB to the AZ scenario is
possible in such interferometers, we have also investigated the effect
of imposing a difference $\phi$ between the order parameter phases of
the two superconductors. The results of these calculations are shown
in Fig.~4b. For the closed structure, in the absence
of magnetic field (solid line), $\delta G_{\text{NSN}}$ increases
monotonically as $\phi$ increases from zero, reaching a maximum at
$\phi = \pi$. The enhancement factor at $\phi=\pi$ is of order
$\sqrt{2}$, which is the ratio of the AZ and BB predictions for UCFs,
thereby demonstrating that a crossover is indeed possible in closed
interferometers. In the presence of a magnetic field (dashed line),
the fluctuations become nearly independent of the phase difference
$\phi$ since time reversal symmetry is already broken by the field. 
Remarkably, the corresponding curve for the open structure shows no
phase dependence. At present, there exists no analytic derivation of
this result. In contrast with the mean conductance $\langle G\rangle$,
which at $E=0$ is independent of $\phi$, but exhibits large-scale
oscillations at finite
$E$~\cite{hartog,petrashovvolkovstoofpothiercourtois}, we have 
found no significant phase dependence in the fluctuations, even at
energies of the order of the Thouless energy.  

Figures 3 and 4 demonstrate that if the N-S interfaces are clean
enough, the magnitude of UCFs are sensitive to the onset of
superconductivity in the S contacts. As a final result, we note that
this effect is suppressed by the presence of Schottky barriers at the
N-S interfaces. This

\begin{figure}[h]
\centering
\epsfig{file=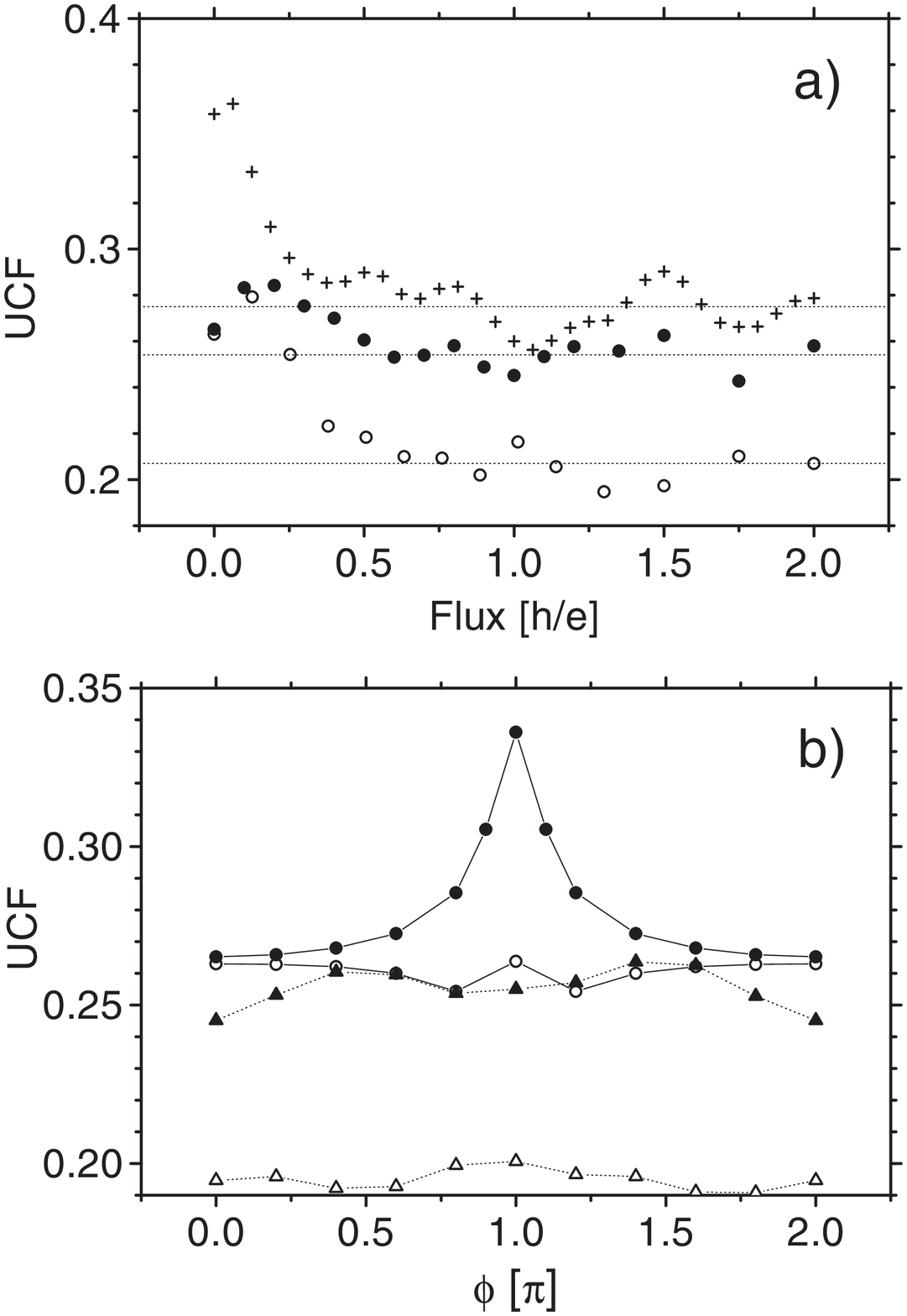,width=8.26cm}
\label{fig:trsphase}
\end{figure}

{\small FIG. 4. a) External magnetic field breaks time reversal
symmetry and decreases the magnitude of fluctuations by a factor of
1.3 both in a normal structure (crosses) and in an open interferometer
(open circles). However, for the closed structure of Fig.~1b (solid 
circles), the suppression is only of the order of 10\%. The dashed
lines at $\delta G_{\text{NSN}}(B \ne 0)= 0.207$, $\delta
G_{\text{NSN}}(B \ne 0)=0.254$ and $\delta G_{\text{NN}}(B \ne
0)=0.275$ show the average of fluctuations for fluxes $\Phi >
\Phi_0/2=h/2e$ in open, closed, and normal structures,
respectively. b) Magnitude $\delta G_{\text{NSN}}$ of universal
conductance fluctuations as functions of the phase difference
$\phi=\phi_1-\phi_2$ between the superconducting segments in the
absence (solid line and circles) and the presence (dashed line
and triangles) of a magnetic field. Results are presented for both
closed (solid markers) and open interferometers (open markers).}

\smallskip

\noindent is demonstrated in Fig.~5 which shows $\delta
G_{\text{NSN}}$ as a function of the barrier conductance $G_B=N\Gamma$
for different numbers of channels $N$ in the contact and
transmissivities $\Gamma$. For $G_B \lesssim 10 e^2/h$, the
fluctuations approach those of a purely normal structure and for $G_B
\gtrsim 20 e^2/h$ fluctuations characteristic of a N-S-N structure
with no barriers are obtained. From an experimental viewpoint, this
suppression could be used as a novel probe of the strength of such
barriers.

In summary, we have used the numerical solution of the Bogoliubov-de
Gennes equations to demonstrate that conductance fluctuations in  
normal-superconducting mesoscopic structures depend on (i) the
presence or absence of quasiparticle transmission through the
structure, (ii) the external magnetic field, and (iii) the
order-para-

\begin{figure}[h]
\centering
\epsfig{file=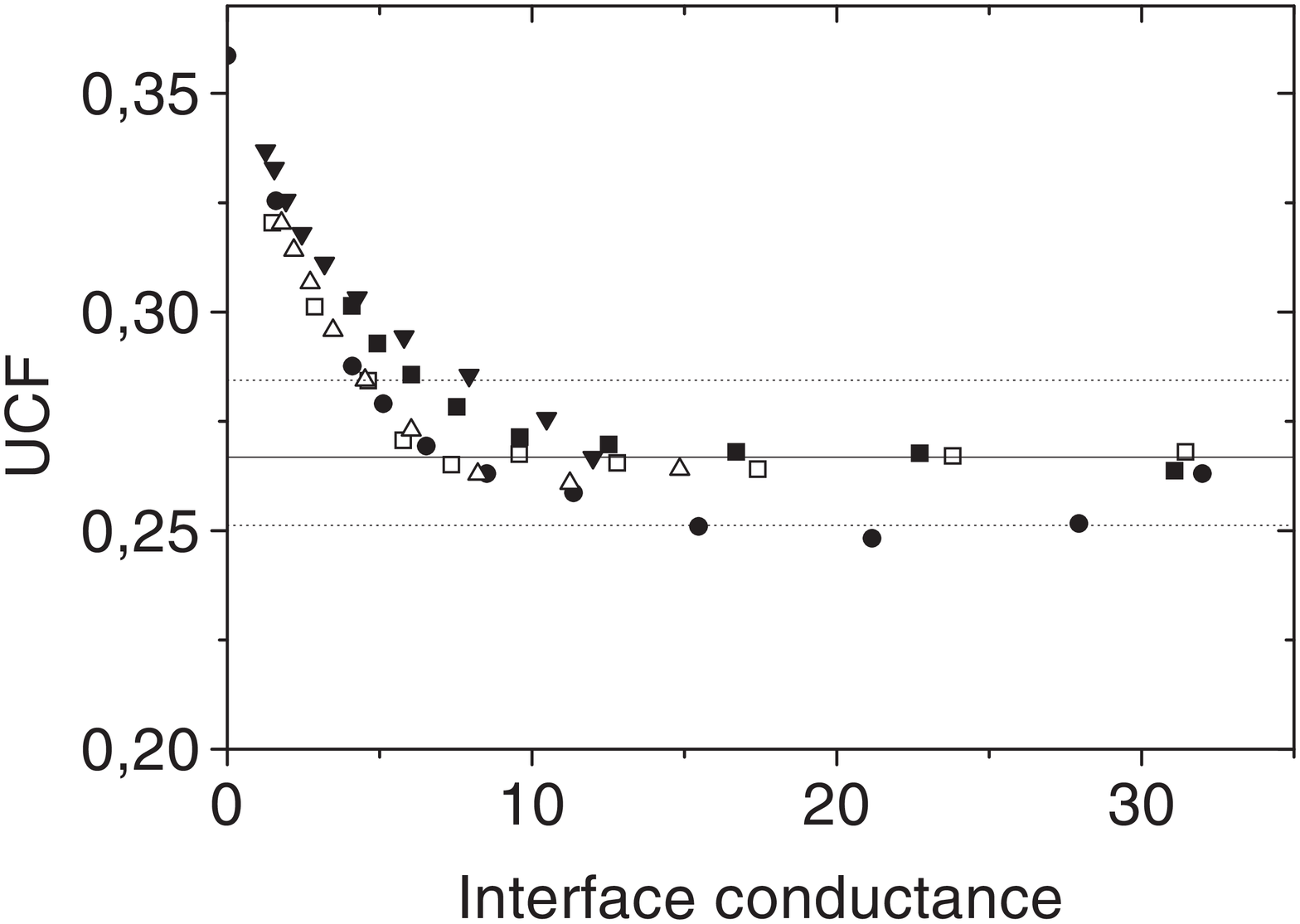,width=8.26cm}
\label{fig:barrier}
\end{figure}

{\small FIG. 5. Tunnel barrier at the N-S interface of the open
structure in Fig.~1a weakens the effect of superconductivity and for
barrier conductance $G_B \protect\lesssim 10e^2/h$, the fluctuations
start to resemble those in the absence of superconductivity. Here the
barrier conductance is expressed in the units of $2e^2/h$ and
fluctuations are plotted for 47 (solid square markers), 38 (open 
squares), 35 (solid circles), 23 (open triangles) and 19 (solid
triangles) open channels at the interface.}

\smallskip

\noindent meter phase difference between the superconducting
contacts. If the phase of the superconducting order parameter is
constant and time-reversal symmetry is not broken by a magnetic field, 
the open and closed structures possess the same UCFs, namely $\delta
G_{\text{NSN}} \approx 0.26 [2e^2/h]$. In contrast to the predictions
of AZ and BB for N-S structures, this is {\it smaller} by roughly a
factor of $\sqrt{2}$ than the normal-state fluctuations $\delta
G_{\text{NN}}$. Applying a magnetic field to the open structure
decreases $\delta G_{\text{NSN}}$ by an additional factor of
$\sqrt{2}$, a result which lies outside current analytic
calculations~\cite{analyticnote}, whereas for the closed structure the
suppression is found to be much weaker (roughly 10 \%), as suggested
by BB. We have also examined the phase dependence of UCFs and shown
that for a closed structure, $\delta G_{\text{NSN}}$ increases by
almost $\sqrt{2}$ as the phase difference between two superconducting
segments is varied from zero to $\pi$, thereby demonstrating a
crossover

\begin{table}[H]
\begin{tabular}{l|c}
System & $\delta G_{\text{NSN}}/[2e^2/h]$
\\\hline 
Normal, ${\cal T}$ & $0.358$\\
Normal, no ${\cal T}$ & $0.275$ \\
Open, $\phi=0$, ${\cal T}$ & $0.266$\\
Open, $\phi=0$, no ${\cal T}$ & $0.207$\\
Open, $\phi=\pi$, ${\cal T}$ & $0.264$\\
Open, $\phi=\pi$, no ${\cal T}$ & $0.201$\\
Closed, $\phi=0$, ${\cal T}$ & $0.273$\\
Closed, $\phi=0$, no ${\cal T}$ & $0.254$\\
Closed, $\phi=\pi$, ${\cal T}$ & $0.336$\\
Closed, $\phi=\pi$, no ${\cal T}$ & $0.255$\\
\end{tabular}
\caption{Calculated universal conductance fluctuations (std($G$)) for
normal structures, open Andreev interferometers and two N-S contacts
in series (``closed''). In the latter, the conductance fluctuations
through a single contact are obtained by a multiplication by
$\sqrt{8}$. ${\cal T}$ denotes time-reversal symmetry. The relative
error of the magnitudes of fluctuations is obtained from 95 \%
confidence intervals for standard deviation and is of the order of 5
\%.}   
\label{tab:ucftable}
\end{table}



\noindent from the BB scenario to the AZ scenario. In contrast for the
open structure, we find that $\delta G_{\text{NSN}}$ is essentially
independent of phase. The calculated magnitudes of the UCFs are
summarized in Table \ref{tab:ucftable}. 

The numerical simulations have been performed in a Cray C94
of the Center for Scientific Computing (CSC, Finland). TH acknowledges
the postgraduate scholarship awarded by Helsinki University of
Technology and the hospitality of the Lancaster University.


\begin{thebibliography}{1}
\bibitem{checknote}
We have numerically checked that increasing these lengths has no
significant effect on the results.
\bibitem{transmissionnote}
In the open structure at zero magnetic field, where the density of
states is suppressed by the proximity effect, the normal transmission
coefficients are less than 10\% of the Andreev reflection coefficients
and the Andreev transmission coefficients are even smaller. For finite
magnetic field or a phase difference of $\phi = \pi$, they become
comparable.
\end{thebibliography}
\end{document}